\documentclass[a4paper]{jpconf}
\usepackage{graphicx}
\usepackage{cite}
\usepackage{amssymb}

\begin{document}
\title{General properties of astrophysical reaction rates in explosive nucleosynthesis}

\author{Thomas Rauscher$^{1,2}$}

\address{$^1$ Department of Physics, University of Basel, 4056 Basel, Switzerland}
\address{$^2$ Institute of Nuclear Research (ATOMKI), H-4001 Debrecen, POB 51, Hungary}

\ead{Thomas.Rauscher@unibas.ch}

\begin{abstract}
Fundamental differences in the prediction of reaction rates with intermediate and heavy target nuclei compared to the ones with light nuclei are discussed, with special emphasis on stellar modifications of the rates. Ground and excited state contributions to the stellar rates are quantified, deriving a linear weighting of excited state contributions despite of a Boltzmann population of the nuclear states. A Coulomb suppression effect of the excited state contributions is identified, acting against the usual $Q$-value rule in some reactions. The proper inclusion of experimental data in revised stellar rates is shown, containing revised uncertainties. An application to the $s$-process shows that the actual uncertainties in the neutron capture rates are larger than would be expected from the experimental errors alone. Sensitivities of reaction rates and cross sections are defined and their application in reaction studies is discussed. The conclusion provides a guide to experiment as well as theory on how to best improve the rates used in astrophysical simulations and how to assess their uncertainties.
\end{abstract}

\section{Introduction}

Explosive nucleosynthesis environments differ from hydrostatic ones by showing higher temperatures and sometimes also much higher proton or neutron densities. As a consequence, the nucleosynthesis processes cannot only extend to nuclei with larger masses than those involved in hydrostatic burning but may also include short-lived isotopes. In addition, the difference between stellar and laboratory reaction rates is more pronounced, due to the higher temperatures encountered as well as the higher average level density in heavier nuclei. This leads to fundamentally different challenges in the prediction and measurement of the involved reaction cross sections and rates than for light nuclei. An overview of some of the relevant effects is given here. For further details, see \cite{review,sensipaper,sprocuncert} and the works cited hereafter.

After introducing the basic definition of the stellar rate in section \ref{sec:defs}, the size of the individual contributions from ground and excited states is discussed in \ref{sec:excstates}. The relation between forward and reverse rates is shown in section \ref{sec:forwrev} and exceptions to the general rules are pointed out in \ref{sec:supp}. The connection between experimental data and stellar rates is studied in section \ref{sec:exp}. Equilibria appearing in high-temperature burning are briefly introduced in section \ref{sec:equilibria} before continuing to the discussion of sensitivities of rates and cross sections to variations in nuclear properties in section \ref{sec:sensi}. The paper is concluded by a brief outline of how to best proceed in selecting nuclides and reactions for future studies.

\section{Stellar rates and impact of thermal excited state population}

\subsection{Basic definition}
\label{sec:defs}

The astrophysical reaction rate $r^*$ for an interaction between two particles or nuclei in a stellar environment is obtained by folding the Maxwell-Boltzmann energy distribution $\Phi$, describing the thermal c.m.\ motion of the interacting nuclei in a plasma of temperature $T$, with the probability $\sigma^*$ that the reaction occurs and by multiplying the result with the number densities $n_a$, $n_A$, i.e., number of interacting particles in a unit volume,
\begin{equation}
\label{eq:rate}
r^*=\frac{n_a n_A}{1+\delta_{aA}} \int_0^\infty \sigma^*(E) \Phi(E,T)\;dE = \frac{n_a n_A}{1+\delta_{aA}} R^* \quad.
\end{equation}

The stellar reactivity (or rate per particle pair) is denoted by $R^*$. To avoid double counting of pairs, the Kronecker symbol $\delta_{aA}$ is introduced. It is unity when the nuclei $a$ and $A$ are the same and zero otherwise. The asterisk superscript indicates \textit{stellar} quantities, i.e., including the effect of thermal population of excited nuclear states in a stellar plasma. Depending on temperature and nuclear level structure, a fraction of nuclei is present in an excited state in the plasma, instead of being in the ground state (g.s.). This has to be considered when calculating the interactions and rates. The population $P_i=p_i/p_0$ of an excited level with spin $J_i$ and energy $E_i$ relative to the g.s.\ with spin $J_0$ is given by \cite{fowler,wardfowler}
\begin{equation}
P_i=\frac{2J_i+1}{2J_0+1}\exp \left(-\frac{E_i}{kT}\right) \quad.
\end{equation}
The stellar cross section $\sigma^*$ appearing in equation (\ref{eq:rate}) can then be shown to be \cite{review}
\begin{eqnarray}
\sigma^*(E,T)&=\frac{\sigma^\mathrm{eff}(E)}{G_0(T)}&=\frac{1}{\sum_i P_i} \sum_i \sum_j \frac{2J_i+1}{2J_0+1} \frac{E-E_i}{E}
\sigma^{i \rightarrow j}(E-E_i) \nonumber \\
&&=\frac{1}{\sum_i P_i} \sum_i \sum_j \frac{2J_i+1}{2J_0+1} W_i \sigma^{i \rightarrow j}(E-E_i)
\quad,
\label{eq:effcs}
\end{eqnarray}
and involves a weighted sum over transitions from all initial excited states $i$ up to the interaction energy $E$, leading to all accessible final states $j$.
As usual, cross sections for individual transitions $\sigma^{i \rightarrow j}$ are zero for negative energies. The quantity $G_0=\sum_i P_i$ is nothing else than the partition function of the target nucleus normalized to the ground state spin and $\sigma^\mathrm{eff}$ is usually called \textit{effective cross section} \cite{holmes}.

\begin{figure}
\begin{center}
\includegraphics[width=\textwidth]{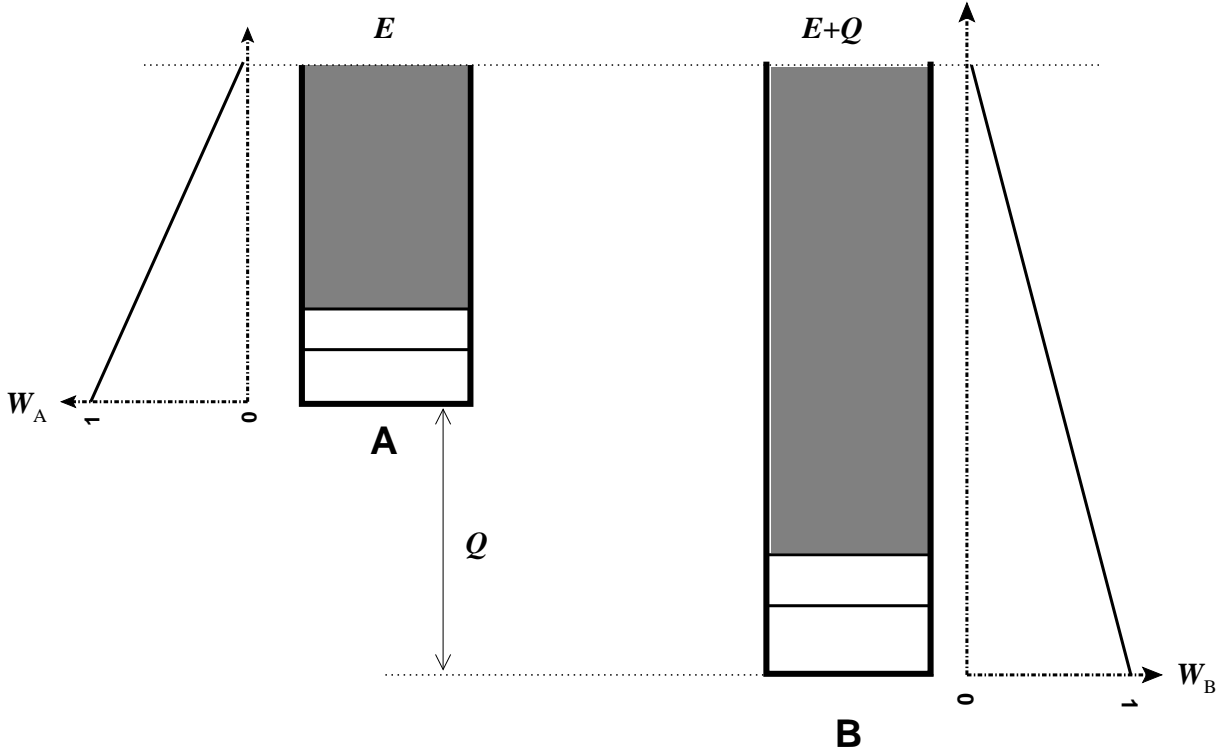}
\caption{\label{fig:weights}Schematic view of the relative effective weights $W_A$ and $W_B$ in the initial and final nucleus, respectively, in the reaction $A(a,b)B$ under the assumption of a positive reaction $Q$-value $Q=Q_{Aa}$.}
\end{center}
\end{figure}

\subsection{Contributions of ground state and excited states to the astrophysical reaction rate}
\label{sec:excstates}

A number of important conclusions can be drawn from equation (\ref{eq:effcs}) \cite{review,sensipaper}. The fact that Maxwell-Boltzmann energy-distributed projectiles act on each excited state leads to a linear energy weighting
\begin{equation}
\label{eq:weights}
W_i=\frac{E-E_i}{E}=1-\frac{E_i}{E}
\end{equation}
of the contributions from excited states in the effective cross section, although their population $p_i=(2J_i+1)\exp \left(-E_i/(kT)\right)$ falls off exponentially with increasing excitation energy $E_i$. The situation is sketched in figure \ref{fig:weights} for a reaction $A(a,b)B$ with positive reaction $Q$-value, where $a$ and $b$ can be a particle or a photon. The relevant energies $E$ are in the energy range contributing most to the integral in equation (\ref{eq:rate}). These energies and the location of the maximum of the integrand can be found in \cite{energywindows}. The simple formula for estimating the Gamow window (see, e.g., \cite{clayton,cauldrons}) from the charges of projectile and target is only applicable when the energy dependence of the cross section is fully given by the entrance channel. This has been found inadequate for many reactions except those involving light nuclei \cite{energywindows,ilibook,newton}.

The linear weighting of the excited state contributions leads to a larger range of excited states contributing to the stellar rate than naively expected from the Boltzmann weight $p_i$. Rates for light nuclei do not exhibit large contributions from excited states and measurements of laboratory cross sections $\sigma_0=\sum_j \sigma^{0 \rightarrow j}$, including only transitions proceeding on the nuclear g.s., are often sufficient to also derive the stellar rate. This is not only due to the larger level spacings compared to heavier nuclei but also because their astrophysical interaction energies are lower in general. The number of possibly contributing levels within the energy range up to $E$ is the determining quantity. On the other hand, rates for intermediate and heavy target nuclei are often dominated by transitions from excited states, both because those nuclei have a larger level density and the relevant interaction energies are also higher, especially for charged particle reactions.

\begin{figure}
\begin{center}
\includegraphics[angle=-90,width=\textwidth]{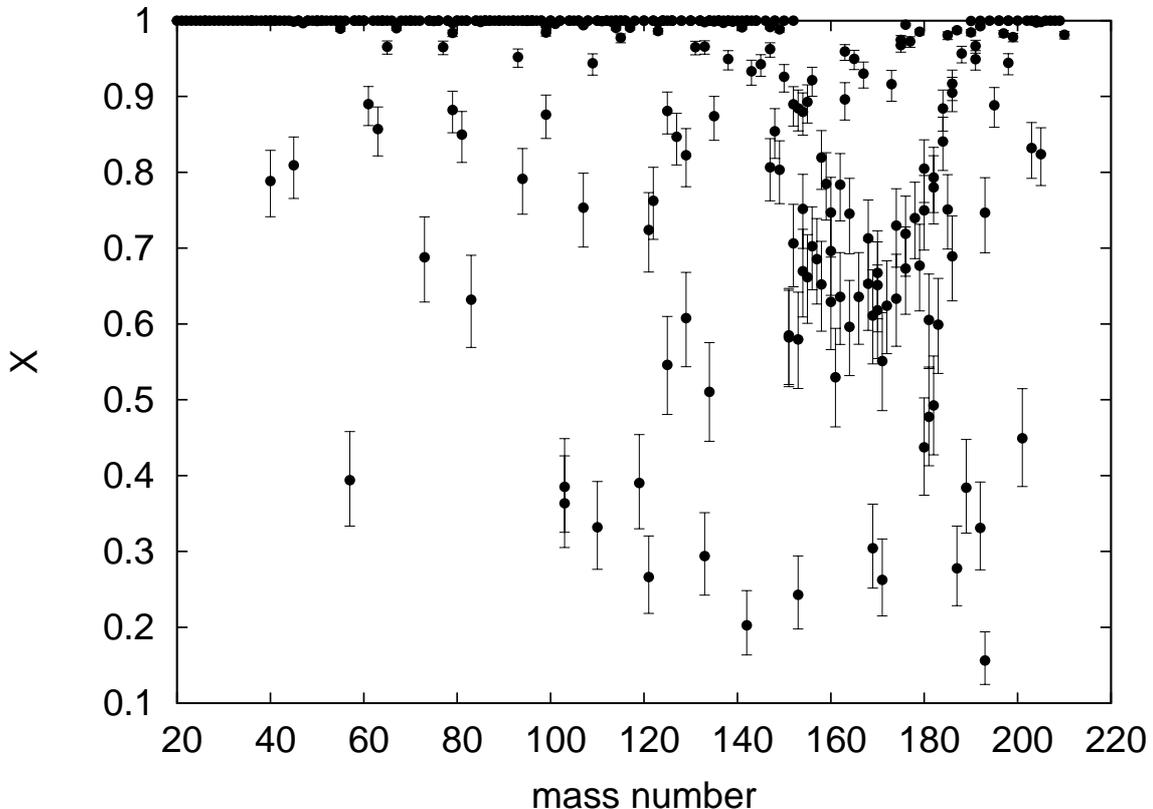}
\caption{Ground state contributions $X=X_0$ to stellar (n,$\gamma$) rates at $kT=30$ keV for all nuclides contained in \cite{kadonis}. The uncertainties were calculated as explained in section \ref{sec:exp}.\label{fig:excstates}}
\end{center}
\end{figure}

\begin{figure}
\begin{center}
\includegraphics[angle=-90,width=\textwidth]{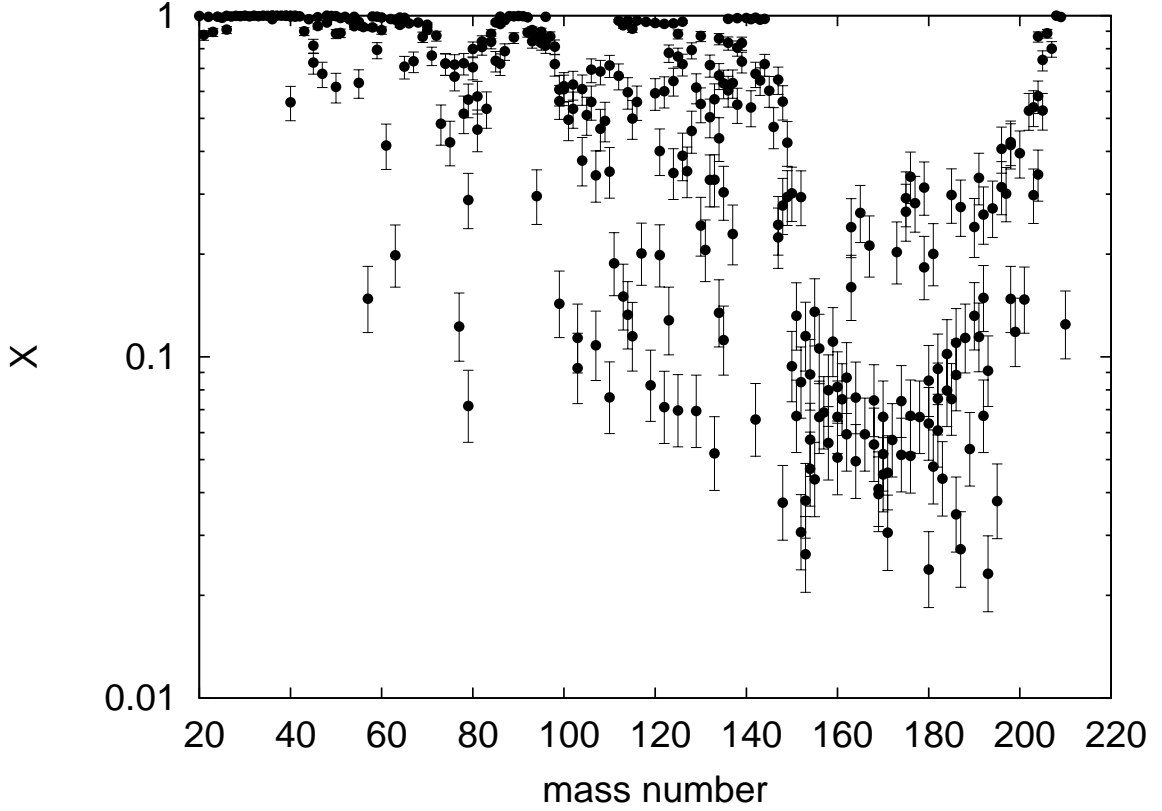}
\caption{\label{fig:2p5comparison}Same as figure \ref{fig:excstates} but for $T=2.5$ GK. Note the logarithmic scale.}
\end{center}
\end{figure}

It is interesting to know the relative contribution $X_i$ of a specific level $i$ to the total stellar rate $r^*$. This is given by \cite{sprocuncert}
\begin{equation}
\label{eq:xfactor}
X_i(T)=\frac{2J_i+1}{2J_0+1}e^{-E_i/(kT)}\frac{\int\sigma_i(E)\Phi(E,T)dE}{\int\sigma^\mathrm{eff}(E)\Phi(E,T)dE} \quad,
\end{equation}
where $\sigma_i=\sum_j \sigma^{i \rightarrow j}$, as before. For the ground state, this simplifies to \cite{xfactor}
\begin{equation}
\label{eq:gsxfactor}
X_0(T)=\frac{\int\sigma_0(E)\Phi(E,T)dE}{\int\sigma^\mathrm{eff}(E)\Phi(E,T)dE} \quad.
\end{equation}
It is very important to note that this is different from the simple ratio $R_0/R^*$ of g.s.\ and stellar reactivity, respectively (see section \ref{sec:exp})!

The relative contribution $X_i$ has several convenient properties. Its upper bound is a value of unity which translates into 100\% contribution to the stellar rate, i.e., the stellar rate is fully determined by the reaction cross section of level $i$. Furthermore, it only assumes values in the range $0\leq X_i\leq 1$. The value of $X_0$ therefore decreases monotonically with increasing plasma temperature $T$. Finally, when including possible uncertainties of reaction model predictions, the uncertainty in $X_i$ scales inversely with the value of $X_i$ but keeping large values of $X_i$ large and small ones small \cite{xfactor} (see also section \ref{sec:exp}).

Complete tables of g.s.\ contributions $X_0$ for reactions on target nuclei between the driplines from $10\leq Z\leq 83$ are given in \cite{sensipaper}. As an example, figure \ref{fig:excstates} shows $X_0$ for (n,$\gamma$) rates close to stability, required for $s$-process studies. Although the typical $s$-process temperature of $kT=30$ keV \cite{kae11} is quite low compared to level spacings, considerable excited state contributions (i.e., $X_0\ll 1$) are found for a surprisingly large number of target nuclei. This is due to the linear fall-off of weights with excitation energy, as discussed above. Naturally, the excited state contributions are largest (i.e., $X_0$ is smallest) in the region of deformed nuclei because they exhibit higher level densities. Figure \ref{fig:2p5comparison} shows $X_0$ for the same nuclei but for a temperature of 2.5 GK, typical for $\gamma$-process nucleosynthesis \cite{arngorp,Rauscher_NIC}. The fast decrease of $X_0$ with increasing temperature can be seen clearly.

Although not discussed in further detail here, it is worth mentioning that also weak interactions and decays are affected by the thermal population of excited states.

\subsection{Comparison between forward and reverse rate}
\label{sec:forwrev}

The well-known reciprocity relation for nuclear reactions shows the relation between the cross section $\sigma^{i \rightarrow j}_{Aa}$ of a reaction proceeding from state $i$ in nucleus $A$ to the final state $j$ in nucleus $B$ and its reverse $\sigma^{j \rightarrow i}_{Bb}$, starting from nucleus $B$, to be \cite{BW52,ilibook}
\begin{equation}
\label{eq:reci_single}
\sigma^{j \rightarrow i}_{Bb}(E_{Bb})=\frac{1+\delta_{Bb}}{1+\delta_{Aa}}\frac{(2J_i+1)(2J_a+1)}{(2J_j+1)(2J_b+1)}\frac{m_{Aa} E_{Aa}}{m_{Bb} E_{Bb}}\sigma^{i \rightarrow j}_{Aa}(E_{Aa}) \quad,
\end{equation}
where $m_{Aa}$, $m_{Bb}$ are the reduced masses and $E_{Aa}$, $E_{Bb}$ the center-of-mass energies relative to the levels $i$, $j$, respectively.\footnote{Throughout the paper it is assumed that $a$, $b$ are light nuclei (or photons) for which excited states do not have to be taken into account.}
If the ejectile $b$ is not a particle but a photon, the reciprocity relation reads
\begin{equation}
\label{eq:reci_single_photon}
\sigma^{j \rightarrow i}_{B\gamma}(E_\gamma)=\frac{1}{1+\delta_{Aa}}\frac{(2J_i+1)(2J_a+1)}{2J_j+1} c^2 \frac{m_{Aa} E_{Aa}}{E_\gamma^2}\sigma^{i \rightarrow j}_{Aa}(E_{Aa}) \quad.
\end{equation}

Using equations (\ref{eq:reci_single}), (\ref{eq:reci_single_photon}) it is straightforward to show that the stellar reactivities defined with the effective cross section -- connecting all initial states to all final states -- also obey reciprocity \cite{review,fowler,holmes}. This only holds when using the stellar and effective cross sections $\sigma^*$ and $\sigma^\mathrm{eff}$, respectively, in the calculation of the reactivity, \textit{not} with cross sections $\sigma_i$ for a single state (and therefore also not for laboratory cross sections $\sigma_0$). The reciprocity relations for a reaction $A(a,b)B$ and its reverse reaction $B(b,a)A$ are \cite{review,holmes,ilibook}
\begin{equation}
\label{eq:revrate}
\frac{R^*_{Bb}}{R^*_{Aa}}=\frac{(2J_0^A+1) (2J_a+1)}{(2J_0^B+1) (2J_b+1)} \frac{G^A_0(T)}{G^B_0(T)} \left( \frac{m_{Aa}}{m_{Bb}}\right) ^{3/2}e^{-Q_{Aa}/(kT)}
\end{equation}
when $a$, $b$ are particles, and
\begin{equation}
\frac{R_\gamma^*}{R^*_{Aa}}=\frac{(2J_0^A+1) (2J_a+1)}{(2J_0^B+1)} \frac{G^A_0(T)}{G^B_0(T)}
\left( \frac{m_{Aa}kT}{2\pi \hbar^2}\right)^{3/2} e^{-Q_{Aa}/(kT)} \label{eq:revphoto}
\end{equation}
when $b$ is a photon. The normalized partition functions $G^A_0$ and $G^B_0$ of the nuclei $A$ and $B$, respectively, are defined as before.
The photodisintegration reactivity
\begin{equation}
\label{eq:photoreac}
R_\gamma^*=\int_0^\infty \sigma_\gamma^*(E) \Phi_\mathrm{Planck}(E,T)\;dE
\end{equation}
includes a stellar photodisintegration cross section $\sigma_\gamma^*(E)$ defined in complete analogy to the stellar cross section $\sigma^*$ in equation (\ref{eq:effcs}).

In relating the capture rate of $A(a,\gamma)B$ to the photodisintegration rate $\lambda_{B\gamma}^*=n_B R_\gamma^*$, however, it has to be assumed that the denominator $\exp(E/(kT))-1$ of the Planck distribution $\Phi_\mathrm{Planck}$ for photons appearing in equation (\ref{eq:photoreac}) can be replaced by $\exp(E/(kT))$, similar to the one of the Maxwell-Boltzmann distribution $\Phi$ with the same temperature $T$. The validity of this approximation has been investigated independently several times \cite{wardfowler,sm144lett,ilibook,review,mathewsreci}.
The contributions to the integral in (\ref{eq:photoreac}) have to be negligible at the low energies where $\Phi$ and $\Phi_\mathrm{Planck}$ differ considerably.
This is assured by either a sufficiently large and positive $Q_{Aa}$, which causes the integration over the Planck distribution to start not at zero energy but rather at a sufficiently large threshold energy, or by vanishing effective cross sections at low energy due to, e.g., a Coulomb barrier. It turns out that the change in the denominator is a good approximation for the calculation of the rate integrals and introduces an error of less than a few percent for astrophysically relevant temperatures and rate values. The assumption may not be valid for s-wave neutron captures with very small (of the order of $Q\lesssim kT$) or negative $Q$-values but the required correction still is only a few \% as can be shown in numerical comparisons between photodisintegration rates calculated with the two versions of the denominator \cite{review}. Generally, larger errors appear at lower temperature. This results in astrophysical irrelevance of the errors in many cases because either the rates are too low (especially for rates involving charged projectiles) or the target nuclei in question are so short-lived that they will never be produced at low plasma temperature. The largest error found was between 50 and 100\% for a few heavy nuclei at the driplines for proton- or $\alpha$-capture at $T<0.3$ GK. For neutron captures, the errors when applying the standard approximation for the reverse rate were never larger than 10\% at any investigated temperature, even at the driplines. For $s$-process neutron captures, the error is completely negligible. It is also negligible for photodisintegrations and their reverse captures in the $\gamma$-process, even though they involve ($\alpha$,$\gamma$) reactions with strongly negative $Q$-values. For more details, see \cite{review}.

The reciprocity relations shown above are very important in reaction networks. Specifying the reactivity of only one reaction direction as input and employing the expressions (\ref{eq:revrate}) and (\ref{eq:revphoto}) to compute the reverse direction avoids numerical inconsistencies which may arise when forward and reverse rates are calculated separately (or even from different sources). The proper balance between the two reaction directions can only be achieved in such a treatment. Furthermore, simplified equations for reaction equilibria (see Sec.~\ref{sec:equilibria}) can be derived which prove important in the modeling and understanding of nucleosynthesis at high temperature.

\begin{table}
\caption{\label{tab:xphoto}Ground state contributions $X_0$ for selected ($\gamma$,n) reactions at 2.5 GK.}
\begin{center}
\begin{tabular}{clclcl}
\br
Target & $X_0$ & Target & $X_0$ & Target & $X_0$ \\
\mr
$^{86}$Sr & 0.00059 &  $^{186}$W & 0.00049 &   $^{198}$Pt & 0.0018\\
$^{90}$Zr & 0.00034 &  $^{185}$Re & 0.00021 &   $^{197}$Au & 0.00035\\
$^{96}$Zr & 0.0061 &   $^{187}$Re & 0.00024 &  $^{196}$Hg & 0.00043\\
$^{94}$Mo & 0.0043 &     $^{186}$Os & 0.00016 &   $^{198}$Hg & 0.00084\\
$^{142}$Nd & 0.0028 &    $^{190}$Pt & 0.000069 &  $^{204}$Hg & 0.0088\\
$^{155}$Gd & 0.0012 &    $^{192}$Pt & 0.00011 &  $^{204}$Pb & 0.0059\\
\br
\end{tabular}
\end{center}
\end{table}

Since the astrophysically relevant energies of the reverse reaction $B(b,a)A$ are related to the ones of the forward reaction by $E_\mathrm{rev}=E+Q$ (see figure \ref{fig:weights}), it is obvious that transitions from excited states contribute more to the stellar rate for reactions with negative $Q$-value. The weights $W_B$ decline more slowly and reach to higher excitation energy than the weights $W_A$ (see equation \ref{eq:weights} and figure \ref{fig:weights}). This is especially pronounced in photodisintegrations. Table \ref{tab:xphoto} gives examples for g.s.\ contributions to ($\gamma$,n) rates of intermediate and heavy target nuclides at a temperature typical for the astrophysical $\gamma$-process, involving the photodisintegration of such nuclei. These numbers have to be compared to the ones for the neutron captures shown in figure \ref{fig:2p5comparison}. It is obvious that they are tiny in comparison.

\subsection{Coulomb suppression of excited state contributions}
\label{sec:supp}

A recently identified effect acts against the general $Q$-value rule for excited state contributions introduced in the previous subsection \cite{coulsupplett,coulsuppprc}. It has to be realized that the relative interaction energies $E_{Aa}=E-E_i$ and $E_{Bb}=E+Q-E_j$ decrease with increasing excitation energies $E_i$, $E_j$. If the cross sections for the individual transitions from the states $i,j$ strongly decrease with lower interaction energy, transitions from higher lying excited states (at larger $E_{i,j}$) cannot contribute much to the stellar rate even when their weights are non-negligible. This is the case for charged particle reactions. When the Coulomb barrier of a reaction with negative $Q$-value is much larger in the entrance channel than in the exit channel (examples for this are charged particles in one channel and neutrons or photons in the other), it may offset the larger energy range and suppress excited state contributions on heavy nuclei.

\begin{figure}
\begin{center}
\includegraphics[angle=-90,width=\textwidth]{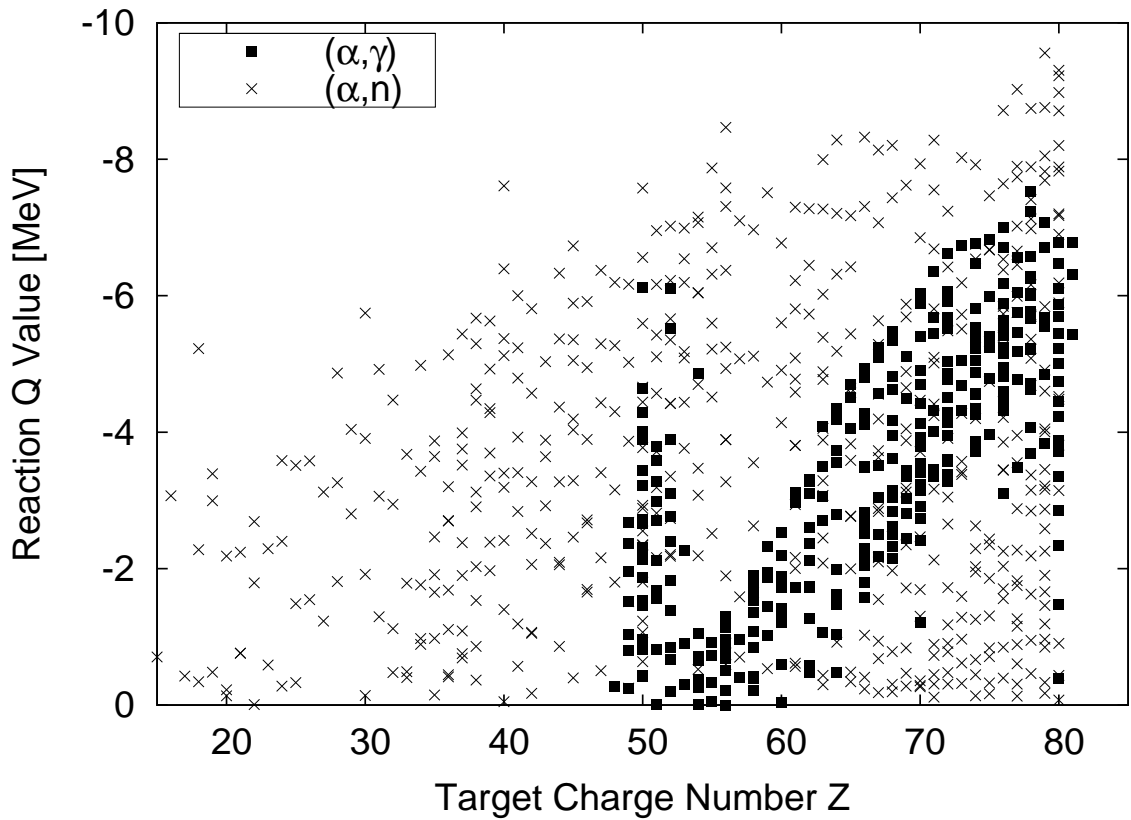}
\caption{Reaction $Q$-values of ($\alpha$,$\gamma$) and ($\alpha$,n) reactions with smaller excited state contribution than those of their reverse reaction despite of $Q<0$, due to the Coulomb suppression effect.\label{fig:qdep}}
\end{center}
\end{figure}

In a large-scale study comparing the contribution of excited states to the stellar rates for forward and reverse reactions\footnote{Note that although the study was using the SEF (see section \ref{sec:exp}) the conclusions still hold.}, more than 1200 such cases were found (which is a small number considering that a total of about 60000 reactions was sampled) \cite{coulsupplett,coulsuppprc}. Not all of them are interesting for astrophysical applications. Noteable cases are ($\alpha$,$\gamma$) reactions at and above Sn, which are important in the $\gamma$-process, and (p,n) reactions on proton-rich nuclei, appearing in the $\gamma$- and $\nu p$-processes \cite{nup}. Figure \ref{fig:qdep} illustrates the Coulomb suppression effect for ($\alpha$,$\gamma$) and ($\alpha$,n) reactions. There is a maximum $|Q|$ in the range of negative $Q$-values appearing in each isotopic chain, indicating how negative the $Q$-value may be while still ensuring that the reaction rate has smaller excited state contributions than those of its reverse reaction. It can clearly be seen that this maximum depends on the charge of the target. The exceptions close to $Z=50$ are due to the low nuclear level densities of even-even nuclides around Sn.

\subsection{Implications for the experimental determination of astrophysical reaction rates}
\label{sec:exp}

Laboratory measurements can only determine the g.s.\ cross section $\sigma_0$ (or the cross section of a long-lived isomeric state). As explained above, this is only sufficient for the determination of the stellar reactivity when the contribution $X_0$ of this level is close to unity. Designing an experiment, it should be taken care to measure a reaction with the largest possible $X_0$. This also implies that the reaction should be measured in the direction of largest $X_0$. As derived above, this usually is the direction of positive $Q$-value with some exceptions. For example, photodisintegration of g.s.\ nuclei always only provides a tiny fraction of the stellar rate (see table \ref{tab:xphoto}) and thus can never be used to constrain a stellar rate for intermediate and heavy nuclei.

The stellar rate can only be derived by combining the experimental data with theory when $X_0<1$. In the past, the stellar enhancement factor (SEF)
\begin{equation}
\label{eq:sef}
f_\mathrm{SEF}=\frac{R^*}{R_0}
\end{equation}
was used. Therein, $R_0$ is the reactivity obtained from folding the experimentally obtained cross section on a g.s.\ (or isomeric state) with a Maxwell-Boltzmann distribution, similar as it is done with the stellar cross section in equation (\ref{eq:rate}). It was pointed out in \cite{xfactor}, however, that $R^*$ and $R_0$ can be similar by chance, even when the $X_0$ are very different. Thus, the SEF is not a suitable measure of the contribution to the stellar rate and it is incorrect to use it for connecting laboratory results to stellar reactivities, although this was done frequently in the past\footnote{Specifically, neutron captures for the $s$-process were incorrectly treated, including the ones in the well-known compilations of \cite{baokaepp,baoetal} and all KaDoNiS versions up to and including v0.3 \cite{kadonis}. The revised table in \cite{sprocuncert} supersedes the stellar rates in \cite{kadonis} and should be used instead.}.

Strictly speaking, the experimental cross section can only replace one of the contributions to the stellar rate while the others remain unconstrained by the data. Knowing $X_0$ and the theory values for $R_0$ and $R^*$, the proper inclusion of a new experimental reactivity $R_0^\mathrm{exp}$ into a new stellar rate is performed as follows. The experimental reactivity cannot be simply multiplied by $f_\mathrm{SEF}$ but rather the \textit{theoretical} stellar reactivity has to be modified to yield the new stellar reactivity, \cite{sprocuncert}
\begin{equation}
R^*_\mathrm{new}= f^* R^* \quad,
\end{equation}
with the renormalization factor
\begin{equation}
f^*=1+X_0\left(\frac{R_0^\mathrm{exp}}{R_0}-1\right) \label{eq:renormstellar}
\end{equation}
containing the experimental result.
Note that the renormalization factor is, of course, temperature-dependent.

Since experimental cross sections are always given with a connected uncertainty (an ``error bar'') and the ultimate goal of a measurement is to reduce the uncertainty inherent in the purely theoretical prediction, it is of particular interest to know the final uncertainty of the new stellar reactivity $R^*_\mathrm{new}$. It is evident that the new uncertainty will only be determined by the experimental one when $X_0$ is large. Otherwise the experiment does not have much impact on the stellar reactivity and also not on its uncertainty. Using $X_0$, also the new uncertainty can be derived. First, the uncertainty of $R_0^\mathrm{exp}$ has to be calculated from the errors in the measured cross sections. It is convenient to use uncertainty factors as is the standard in astrophysical investigations. An uncertainty factor $U_{\mathrm{exp}}\geq 1$ implies that the ``true'' value of $R_0^\mathrm{exp}$ is in the range
$R_0^\mathrm{exp}/U_{\mathrm{exp}}\leq R_0^{\mathrm{true}}\leq R_0^\mathrm{exp}U_{\mathrm{exp}}$. For example, an uncertainty of 20\% translates into $U_{\mathrm{exp}}=1.2$. Secondly, the theoretical uncertainty factor $U_{\mathrm{th}}=U^*$ of the stellar reactivity $R^*$ has to be estimated. The fundamental differences between error determinations in experiment and theory are discussed in \cite{sensipaper} and appropriate choices are suggested in \cite{sprocuncert}. Combining these errors leads to a new uncertainty factor
\begin{equation}
U^*_\mathrm{new}=U_{\mathrm{exp}}+(U^*-U_{\mathrm{exp}})(1-X_0) \label{eq:uncertainty}
\end{equation}
for $R^*_\mathrm{new}$. Here, $U_{\mathrm{exp}}\leq U^*$ is assumed because otherwise the measurement would not provide an improvement.
Obviously, also the uncertainty factor is temperature-dependent because at least $X_0$ depends on the plasma temperature.
It is further possible to consider the uncertainty of $X_0$ in $U^*_\mathrm{new}$. Its impact, however,
is small with respect to the other experimental and theoretical uncertainties. It was shown in \cite{xfactor} that the magnitude of the error scales
inversely proportionally with the value of $X_0$ (or generally $X_i$), i.e., $X_0=1$ has
zero error as long as $G_0$ is known (this is the case close to stability),
and that the uncertainty factor $U_X\geq 1$ of $X_0$ is given by $\max(u_X,1/u_X)$, where
\begin{equation}
u_X=\overline{u}\left(1-X_0\right) + X_0
\end{equation}
and $\overline{u}$ is an averaged uncertainty factor in the predicted \textit{ratios} of the $R_i$. These ratios are believed to be predicted with better accuracy than the rates themselves and so it can be assumed $\overline{u}\leq U_\mathrm{th}$.
In any case, the uncertainties are sufficiently small to preserve the magnitude
of $X_0$, i.e., small $X_0$ remain small within errors and large $X_0$ remain large, as can also be seen in figures \ref{fig:excstates}, \ref{fig:2p5comparison}.

\begin{figure}
\begin{center}
\includegraphics[angle=-90,width=\textwidth]{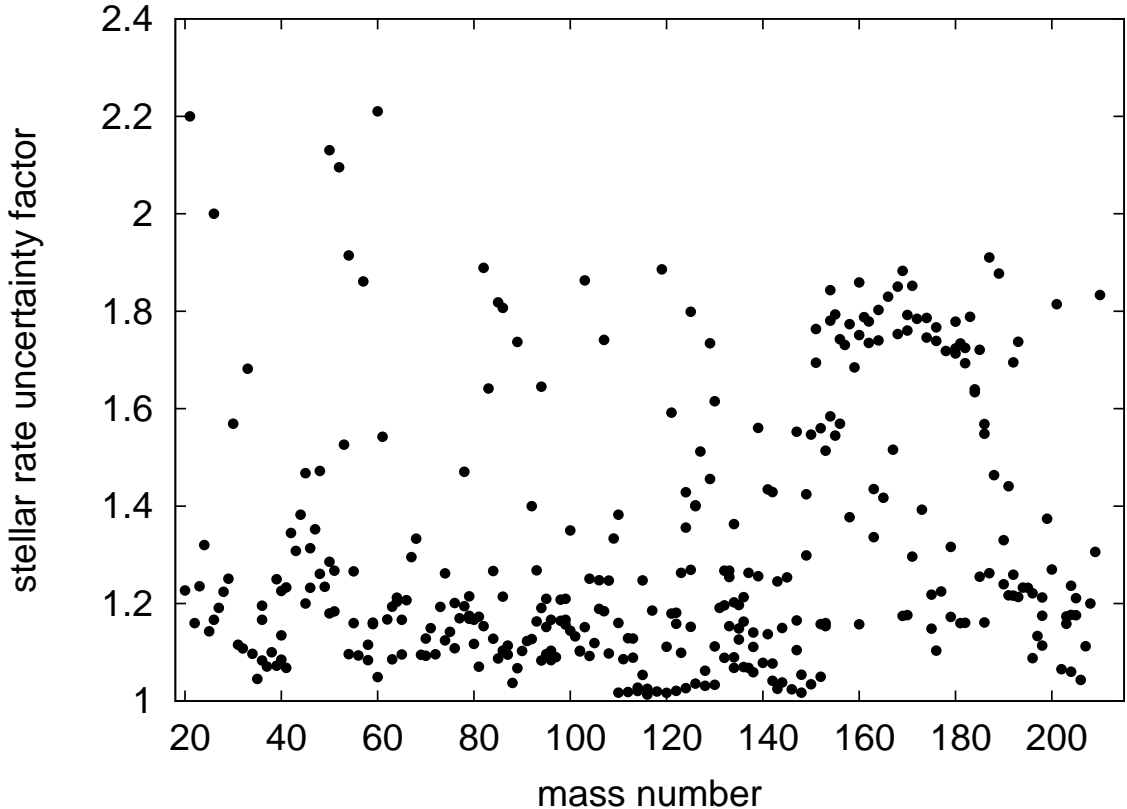}
\caption{Uncertainty factors $U^*_\mathrm{new}$ for stellar (n,$\gamma$) rates at $kT=30$ keV along stability, also including the uncertainty in $X_0$. All nuclides contained in \cite{kadonis} are shown.\label{fig:uncert}}
\end{center}
\end{figure}

It is sometimes stated that the $s$-process is the nucleosynthesis process with the best constrained nuclear input (see, e.g., \cite{kae11,sreview}). This statement, however, is based on the uncertainties in the measured (n,$\gamma$) cross sections. The resulting uncertainties $U^*_\mathrm{new}$ in the stellar reactivities and rates are much larger in many cases \cite{sprocuncert}. Figure \ref{fig:uncert} shows $U^*_\mathrm{new}$ for stellar neutron capture rates at a typical $s$-process temperature for all data contained in \cite{kadonis}, assuming $U_\mathrm{th}=2$ (except for theoretical entries for which a larger uncertainty is given in the compilation). Astrophysical $s$-process studies should consider that the rates can vary within the shown ranges and not only within the experimental errors.

\section{Equilibria}
\label{sec:equilibria}

Experimental and theoretical investigations aiming at improved reaction rates for astrophysics should keep in mind that sometimes it is not necessary to know the rates for calculating the nucleosynthesis products. Forward and reverse reactions occur simultaneously in a plasma. If the reactions in both directions are faster than the nucleosynthesis timescale (the duration of the nuclear burning), then the number of nuclei of a given isotope will saturate to its equilibrium number. This number can be calculated using the reciprocity relations for stellar reactivities shown in section \ref{sec:forwrev}.
Using equations (\ref{eq:revrate}), (\ref{eq:revphoto}) it is trivial to show that
\begin{equation}
\frac{n_A n_a}{n_B n_b}=\frac{(2J_0^A+1) (2J_a+1)}{(2J_0^B+1) (2J_b+1)} \frac{G^A_0}{G^B_0} \left( \frac{m_{Aa}}{m_{Bb}}\right) ^{3/2}e^{-Q_{Aa}/(kT)}
\end{equation}
when $a$,$b$ are particles, and
\begin{equation}
\frac{n_A n_a}{n_B}=\frac{(2J_0^A+1) (2J_a+1)}{(2J_0^B+1)} \frac{G^A_0}{G^B_0} \left( \frac{m_{Aa}kT}{2\pi \hbar^2}\right)^{3/2} e^{-Q_{Aa}/(kT)}
\end{equation}
for a reaction $A(a,\gamma)B$. The individual rates do not appear anymore in the relation between the number densities. Note that this does not imply that the abundances remain constant, they still depend on $T$ which may vary with time, as well as on $n_a$ and $n_b$.

Depending on the plasma density, above $T\approx 4-5$ GK all reactions (with the exception of the weak interaction) achieve equilibrium. It can be shown that the equilibrium number density of a nucleus $A$ can be calculated from a set of three equations \cite{review,arnett,pad}
\begin{eqnarray}
n_A &=&G_A
\frac{(Z+N)^{3/2}}{2^{Z+N}}
\left( \frac{2\pi\hbar^2}{m_u kT} \right)^{Z+N-1} e^{B_A/(kT)}n_\mathrm{n}^{N}n_\mathrm{p}^{Z} \quad, \label{eq:nse} \\
\frac{\rho}{m_u} &=& \sum_A n_A (Z+N) \quad, \label{eq:masscons} \\
Y_\mathrm{e}\left( \frac{\rho}{m_u} \right) &=& \sum_A n_A Z \quad, \label{eq:chargeconv}
\end{eqnarray}
where $Z$, $N$ are the charge and neutron number, respectively, of nucleus $A$, $G_A=(2J_0^A+1)G_0^A$ is its partition function, $n_\mathrm{n}$, $n_\mathrm{p}$ are the number densities of the free neutrons and protons, respectively, and $m_u$ the nuclear mass unit. Nucleus $A$ has a nuclear binding energy of $B_A$. The sums run over all species of nuclei in the plasma, including neutrons and protons. Equation (\ref{eq:masscons}) expresses mass conservation and (\ref{eq:chargeconv}) is the charge conservation. The plasma density is denoted by $\rho$, and $Y_\mathrm{e}$ is the mean number of electrons per baryon. Note that reactions mediated by the weak interaction are not included in the equilibrium and therefore $Y_\mathrm{e}$ may be time-dependent. Again, individual rates do not appear anymore in the above equations.

When all abundances in the network obey the above relations, full \textit{nuclear statistical equilibrium} (NSE) is achieved. In this case, no reaction rates have to be known. The relevant nuclear physics input is the binding energies and the weak interaction rates determining $Y_\mathrm{e}$. In realistic cases, more or less extended groups of nuclei are in statistical equilibrium and the relative abundances within a group can be described by equations similar to (\ref{eq:nse}). The different groups are connected by comparatively slow reactions not being in equilibrium, which determine the abundance level of one group with respect to another group. The rates of these slow, connecting reactions have to be known explicitly. This is called \textit{quasi-statistical equilibrium} (QSE). It appears in various kinds of high-temperature burning, such as hydrostatic oxygen and silicon burning in massive stars \cite{hix} and different explosive scenarios. Proton captures are in equilibrium with ($\gamma$,p) reactions within isotonic chains in the $\nu p$- and $rp$-processes \cite{froh,nup,rp}. The slow reactions connecting different groups (chains) are $\beta^+$-decays and electron captures in the $rp$-process and (n,p) reactions in the $\nu p$-process. On the neutron-rich side of stability, the (n,$\gamma$)-($\gamma$,n) equilibrium of the $r$-process \cite{ctt,arngorr} is a QSE within isotopic chains and slower $\beta^-$-decays connect the chains \cite{dowehavetoknow}. Of course, individual rates of all involved reactions have to be known in order to determine under which conditions (temperature, density) the nuclei fall out of NSE or QSE.

Another situation, in which not all individual rates have to be known, is a sequence of reactions in which most of the reactions are much faster than one or a few slow ones. In this case, the matter processing along this chain is determined by the slow reactions and their reaction rates have the largest impact. When the process has lasted for a sufficient time to move through all the nuclei in a chain and the relative number of nuclei in the chain does not change anymore, this is called \textit{steady flow} \cite{review}. Steady flow considerations are helpful when
investigating hydrostatic hydrogen burning of stars through the pp-chains and the CNO cycles \cite{ilibook,clayton}. In the past they have also been used for sequences of neutron captures in the $s$-process on nuclei in between magic numbers \cite{clayton}.

\section{Sensitivities}
\label{sec:sensi}

The study of the sensitivities of reaction cross sections and astrophysical reactivities to variations in nuclear properties has proven essential in several respects. Firstly, it is not always possible to cover the astrophysically relevant energy range in experimental investigations, especially for charged particle reactions on intermediate and heavy nuclei which show tiny reaction cross sections at low energy due to the high Coulomb barriers. Astrophysical reactivities as well as low-energy cross sections may have quite different sensitivities than cross section at higher energy. Therefore great care has to be taken in comparisons of theory to experiment and in subsequently drawing conclusions on the ability of theory to predict astrophysical reaction rates. Deviations from predictions found in experimental data above the astrophysical energies often turn out to be irrelevant for the rate predictions because they are caused by an insufficient theoretical description of a nuclear property which has no impact at low energy. On the other hand, they may also not be sufficient to constrain the actually relevant properties and thus add nothing to better constraining the reaction rate. In order to draw valid conclusions for astrophysics, experimental and theoretical studies of reaction cross sections have to consider the applicable sensitivities.

As laid out in \cite{sensipaper}, the knowledge of sensitivities also allows to study uncertainties in the prediction of reaction rates. While the definition of an ``error bar'' for theory is complicated by some fundamental differences to attaching an experimental error (see \cite{sensipaper} for details), properly defined sensitivities immediately allow to see the impact of various uncertainties in nuclear properties and input without having to perform the full variational calculations. This also implies that it is easy to see which properties are in need of a better description in order to better constrain the astrophysical rate.

In order to quantify the impact of a variation of a model quantity $q$ (directly taken from input or derived from it) on the final result $\Omega$ (which is either a cross section or a reactivity), the relative sensitivity $s\left(\Omega,q\right)$ is defined as \cite{sensipaper,review}
\begin{equation}
\label{eq:sensi}
s\left(\Omega,q\right)=\frac{v_\Omega-1}{v_q-1} \quad.
\end{equation}
It is a measure of a change by a factor of $v_\Omega=\Omega_\mathrm{new}/\Omega_\mathrm{old}$ in $\Omega$ as the result of a change in the quantity $q$ by the factor $v_q=q_\mathrm{new}/q_\mathrm{old}$, with $s=0$ when no change occurs and $s=1$ when the final result changes by the same factor as used in the variation of $q$, i.e., $s\left(\Omega,q\right)=1$ implies $v_\Omega=v_q$.
Further information is encoded in the sign of the sensitivity $s$. Since both $v_\Omega>0$ and $v_q>0$ for the quantities studied in this context, a positive sign implies that $\Omega$ changes in the same manner as $q$, i.e., $\Omega$ becomes larger when the value of the quantity $q$ is increased. The opposite is true for $s<0$, i.e., $\Omega$ decreases with an increase of $q$.
The above definitions are consistent with the ones used in standard sensitivity analysis when realizing that (\ref{eq:sensi}) is equivalent to
\begin{equation}
s\left(\Omega,q\right)=\frac{\frac{d\Omega}{\Omega_\mathrm{old}}}{\frac{dq}{q_\mathrm{old}}} \quad,
\end{equation}
with $d\Omega=\Omega_\mathrm{new}-\Omega_\mathrm{old}$ and $dq=q_\mathrm{new}-q_\mathrm{old}$.

The varied quantities $q$ in reaction rate studies are neutron-, proton-, $\alpha$-, and $\gamma$-widths. Sometimes also the nuclear level density is varied although it can be shown that it mainly affects the $\gamma$-width in astrophysical applications. Comparing the sensitivities of a cross section to variations of the $\gamma$-widths and level density allows to further constrain an observed deviation of a prediction from data, provided the cross section is mainly sensitive to the $\gamma$-width at the measured energy. As mentioned before, also the impact of uncertainties in the (input) quantities on the final rate or cross section can be studied using the relative sensitivities. When a quantity $q$ has an uncertainty factor $U_q$, it will appear as an uncertainty factor $^\Omega U=\left|s\left(\Omega,q\right) \right| U_q$ in the final result.

\begin{figure}
\begin{center}
\includegraphics[angle=-90,width=\columnwidth]{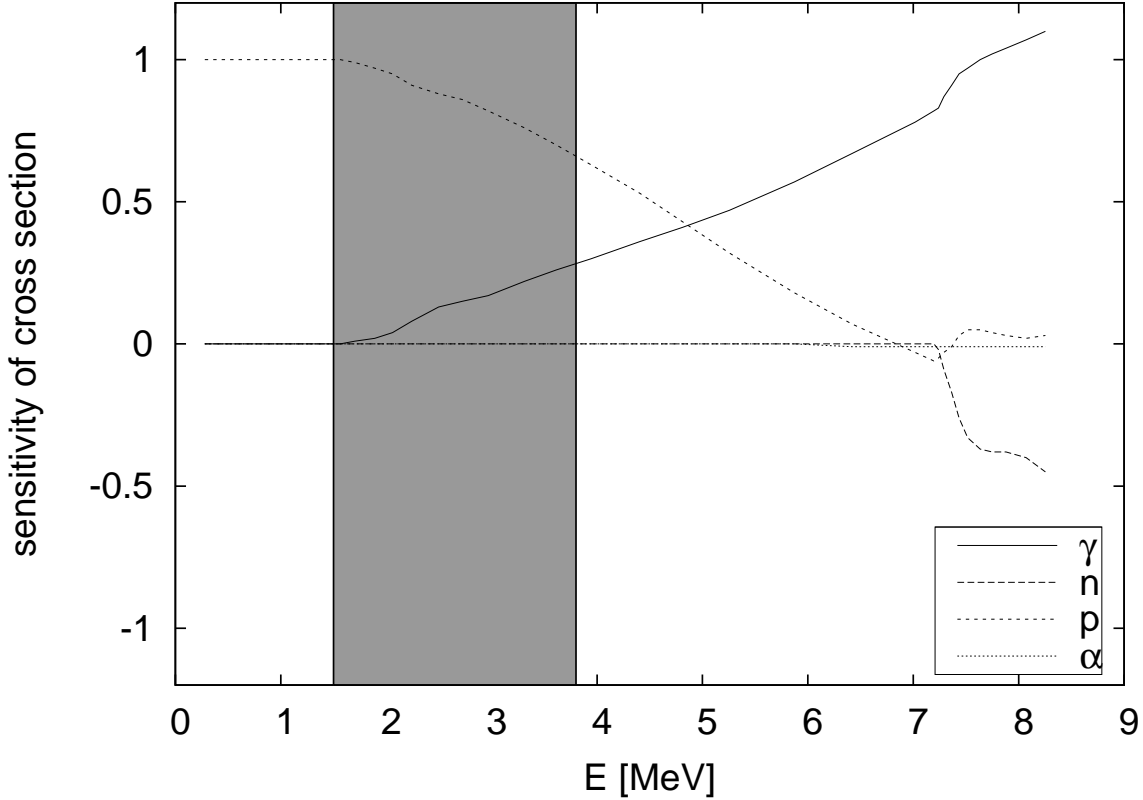}
\caption{Sensitivities $s$ of the laboratory cross section of $^{96}$Ru(p,$\gamma$)$^{97}$Rh to variations of nucleon-, $\alpha$-, and $\gamma$-widths, plotted as functions of c.m.\ energy \cite{sensipaper}. The shaded region is the astrophysically relevant energy range for $2\leq T \leq 3$ GK \cite{energywindows}.\label{fig:ruplot}}
\end{center}
\end{figure}

It should be noted that laboratory cross sections $\sigma_0$ exhibit different sensitivities in forward and reverse reactions, whereas it follows from the reciprocity relations (\ref{eq:revrate}), (\ref{eq:revphoto}) that the sensitivities of the reaction rates are the same for forward and reverse reaction. It is also to be mentioned that even in the astrophysical energy range laboratory cross sections do not necessarily show the same sensitivity as the stellar rate. This depends on $X_0$ and/or the sensitivities of the transitions from excited states. Usually the cross section in the reaction direction with larger $X_0$ will behave more similar to the stellar rate.

Extended tables of sensitivities for reactions on target nuclei between the driplines and with $10\leq Z\leq 83$ have been published in \cite{sensipaper}, for rates as well as cross sections. Here, only a few examples are discussed. Figure \ref{fig:ruplot} shows a striking example of how different sensitivities can be in the astrophysical energy range and above it. If a measurement of $^{96}$Ru(p,$\gamma$)$^{97}$Rh showed discrepancies to predictions above, say, $5-6$ MeV it would be hard to disentangle the uncertainties from different sources. Moreover, such a discrepancy would say nothing about the quality of the prediction at astrophysical energies, as the cross section is almost exclusively sensitive to the proton width there, whereas the proton width does not play a role at higher energy. Obviously, if good agreement is found, on the other hand, between experiment and theory at higher energy, this does not constrain the uncertainty of the reaction rate.

As a general observation it was found that the $\gamma$-width is not relevant in astrophysical charged particle captures on intermediate and heavy mass nuclei \cite{sensipaper}. This is due to the fact that the Coulomb barrier suppresses the proton- and $\alpha$-widths at low energy and makes them smaller than the $\gamma$-widths at astrophysical interaction energies. The cross sections and rates will always be most sensitive to the channel with the smallest width. It also follows from this that whenever an $\alpha$-particle is involved, the reaction will be mostly sensitive to its channel. The situation for neutron captures is more diverse. For neutron-rich nuclei the rates are strongly sensitive to the $\gamma$-width. In between magic neutron numbers along stability and especially in the region of deformed nuclei, however, they are also or even more sensitive to the neutron-width (see figures 14, 15 in \cite{sensipaper}). This can be explained by the fact that the compound nuclei in these regions have higher level densities and this leads to comparable sizes of the neutron- and $\gamma$-widths at astrophysical energies.

Further examples regarding the application of sensitivity plots to the astrophysical interpretation of experimental data are found in \cite{gyurky06,gyurky07,kiss07,gyurky10,review,sauer11,hal12,rautm169,sauer12} and references therein.

\section{Conclusion}

While reactions involving light target nuclei are dominated by few, easily identifyable transitions, the dependence of astrophysical reaction rates and cross sections is more complicated for intermediate and heavy mass nuclei. An overview of the important effects was given above. Suggestions on how to implement the above results to improve our knowledge of astrophysical reaction rates have been given in \cite{review,sensipaper,sprocuncert}.

Crucial for all experimental and theoretical investigations is the choice of nuclei, reactions, and nuclear properties of astrophysical relevance to study. This involves several considerations. Firstly, not all nuclei and reactions appearing in a nucleosynthesis process are equally important. Some may be rendered insignificant by full or partial equilibria (section \ref{sec:equilibria}). Furthermore, the g.s. contributions $X_0$ to the stellar rate should also be considered when devising an experiment to judge whether a feasible measurement can actually constrain the stellar reaction rate, even when performed within the astrophysical energy range. A reaction should preferrably be measured in the direction of largest $X_0$. The applicable energy ranges can be found in \cite{energywindows}. If a measurement within this energy range is possible, the theoretical rate has to be corrected by the experimental data as explained in section \ref{sec:exp}. This also assigns an improved uncertainty. A simple renormalization is not adequate, not even when $X_0=1$ at one temperature because of the temperature-dependence of $X_i$. If a measurement is unfeasible at astrophysical energies, the sensitivities introduced in section \ref{sec:sensi} have to be consulted to assess whether cross sections in the accessible energy range can help to constrain the rate. Knowing the sensitivities, also alternative reactions or approaches may be sought to extract the actually important quantities required as input for the rate predictions.

All in all, the procedures laid out here allow to focus experiment and theory on the study of the actually relevant nuclear reactions and properties. They provide a well-directed approach to a purposeful improvement of astrophysical reaction rates, necessary for the best and most economical use of existing and future experimental and computing facilities.

\newpage

\ack
This work was supported in part by the European Commission within the FP7 ENSAR/THEXO project and the EuroGENESIS Collaborative Research Programme. TR also acknowledges support through a "Distinguished Guest Scientist Fellowship" from the Hungarian Academy of Sciences.

\end{document}